\documentstyle[prl,aps,floats,epsf,color]{revtex}
\begin{document}
\baselineskip=12pt
\def\be{\begin{equation}}
\def\ee{\end{equation}}
\def\bea{\begin{eqnarray}}
\def\eea{\end{eqnarray}}
\def\E{{\rm e}}
\def\bearst{\begin{eqnarray*}}
\def\eearst{\end{eqnarray*}}
\def\peleven{\parbox{11cm}}
\def\peffec{\peight{\bearst\eearst}\hfill\peleven}
\def\pspace{\peight{\bearst\eearst}\hfill}
\def\ptwelve{\parbox{12cm}}
\def\peight{\parbox{8mm}}
\twocolumn[\hsize\textwidth\columnwidth\hsize\csname@twocolumnfalse\endcsname

\title
{Height Fluctuations and Intermittency of $V_2 O_5$ Films by
Atomic Force Microscopy }

\author
{A. Iraji zad $^{a}$, G. Kavei $^{b}$, M. Reza Rahimi Tabar
$^{a,c,d}$,
\\
and S.M. Vaez Allaei $^{e}$}
\address
{\it $^a$ Dept. of Physics, Sharif University of Technology,
P.O.Box 11365-9161, Tehran, Iran\\
$^b$ Material and Energy, Research Center, P. O. Box 14155-4777, Tehran, Iran \\
$^c$ CNRS UMR 6529, Observatoire de la C$\hat o$te d'Azur,
BP 4229, 06304 Nice Cedex 4, France\\
$^d$ Dept. of Physics, Iran University of Science and Technology,
Narmak, Tehran 16844, Iran\\
$^e$ Institute for Advanced Studies in Basic Sciences, P. O. Box
45195-159, Zanjan , Iran
\\
}

\maketitle


\begin{abstract}

The spatial scaling law and intermittency of the $V_2 O_5$
surface roughness by atomic force microscopy has been
investigated. The intermittency of the height fluctuations has
been checked by two different methods, first, by measuring
scaling exponent of q-th moment of height-difference fluctuations
i.e. $C_q = < |h(x_1) - h(x_2)|^{q}
>$ and the second, by defining generating function $Z(q,N)$ and generalized
multi-fractal dimension $D_q$. These methods predict that there is
no intermittency in the height fluctuations. The observed
roughness and dynamical exponents can be explained by the
numerical simulation on the basis of forced Kuramoto-Sivashinsky
equation.

\end{abstract}
\hspace{.3in}
\newpage
]
\section{Introduction}
 Due to the technical importance and fundamental interest, a
great deal of effort has been devoted to understanding the
mechanism of thin-film growth and the kinetic roughening of
growing surfaces in various growth techniques. Analytical and
numerical treatments of simple growth models suggest, quite
generally, the height fluctuations have a self-similar character
and their average correlations exhibit a dynamic scaling form
[1-6].

 Vanadium pentoxide, $V_2 O_5$, has been the subject of intense
work because of its diverse applications in catalytic oxidation
reactions, cathodic electrode in solid state micro-batteries,
windows and electrochromic devices as well as gas sensors and to
be of interest for transmittance modulation in
  smart windows with potential application in the architecture and
  automotive. Also $V_2 O_5$ is a low mobility semiconductor,
  and having predominantly an n-type. Electrons are the charge
  carriers, and an increase in the carrier density in accompanied
  by reduction in oxygen concentration in the lattice [7-9].

This work aims to study the roughness and dynamical exponents ( $
\chi$ and $z$)  and the intermittency  of the $V_2 O_5$ films by
the atomic force microscopy. We measure the height-difference
moments $C_q (l=|x_1 -x_2|) = < |h(x_1) - h(x_2) |^q
> $ and show that they behave as $ \sim |x_1 - x_2 |^ {\xi _q}$.
 The obtained $\xi_q$ is a linear function of $q$. We also
 introduce the generating function $Z(q,N)$ and generalized
multifractal dimension $D_q$ and show that $D_q$ behave also as a
linear function of $q$. These observations indicate that the
height fluctuations are not intermittent.
 It is also argued that the measured roughness and dynamical exponents
 belong to
  early-time scaling regime of the noisy Kuramoto-Sivashinsky
 equation.

\section{Experiments }

$V_2O_5$  layers were grown on the polished Si(100) substrate by
resistive evaporation method in a high vacuum chamber.
The pressure during evaporation was $10^{-5}$ torr.
The
thickness of the growing films was measured in situ by a quartz
crystal thickness monitor.
We performed all deposition at room teperature,
with deposition rate about $10-15 nm/min$.
However, during the film deposition the substrate temperature raised to $T \sim 60 $$C^0$, due to
the radiation effect of alumina boat. The substrate temperature was determined using
chromel/alumel thermocouple mounted in close proximity of samples.
 Surface composition of samples was
measured by Auger electron spectroscopy (AES) using a $3$ keV
electron beam and a cylindrical mirror analyzer (Varian model
981-2607). The surface topography of the films was investigated
using Park Scientific Instruments model Autoprobe CP. The images
are collected in a constant force mode and digitized into $256
\times 256 $ pixels with scanning frequency of $0.6$ Hz. The
cantilever of 0.05 N $m^{-1}$ spring constant with a commercial
standard pyramidal $Si_3N_4$ tips has been used. A variety of
scans, each with size $L$ were recorded at random locations on the
$V_2O_5$ film surface.
  In order to determine the structure of the deposited $V_2O_5$ films,
we have performed XRD measurements of samples. The spectra for the
as-deposited $V_2O_5$ films grown showed that the $V_2O_5$ thin
films are amorphous.
 For thick films ($d>200 nm$) two very broad weak peaks were observed representing the
growth of  (001) and (002) peaks of the $V_2O_5$  orthorombic
structure[7]. The yellow color of deposited oxide films and their
optical transmission spectra indicates the existence of $V_2O_5$
structure rather than other vanadium oxide phases[8]. AES analysis
of the $V_2O_5$ samples showed $V$ and $O$ peaks at the surface of
the deposited films.
 The stoichiometry of the
vanadium oxide films was calculated from the ratio of $O$ to $V$
Auger peak heights by considering the elemental sensitivity
factors. The $O/V$ ratio was $2.5 \pm 0.1$ indicating the
formation of stoichiometric vanadium pentoxide at the surface of
the thin film ( see Fig.1). With these observations we conclude
the film composition is nearly stiochiometric[9].

\begin{figure}
\epsfxsize=7.9truecm \epsfbox{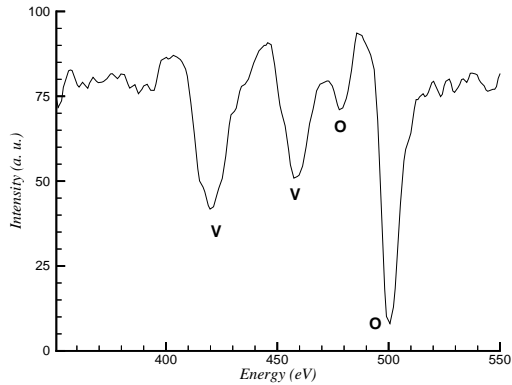} \narrowtext \caption{ AES
spectra from $V_2 O_5$ surface. The $O/V$ ratio is $2.5 \pm 0.1$ (
by considering the elemental sensitivity factors). }
\end{figure}

\section{ Results }
Figure 2 shows the variation of the surface morphology for
different growth times ( or thickness) of processed samples by
AFM. As deposition proceeds, the size of mountains and valleys
grows until the system approaches to a stationary state in which
the thickness is about $150 nm$.

\begin{figure}
\epsfxsize=7.4truecm\epsfbox{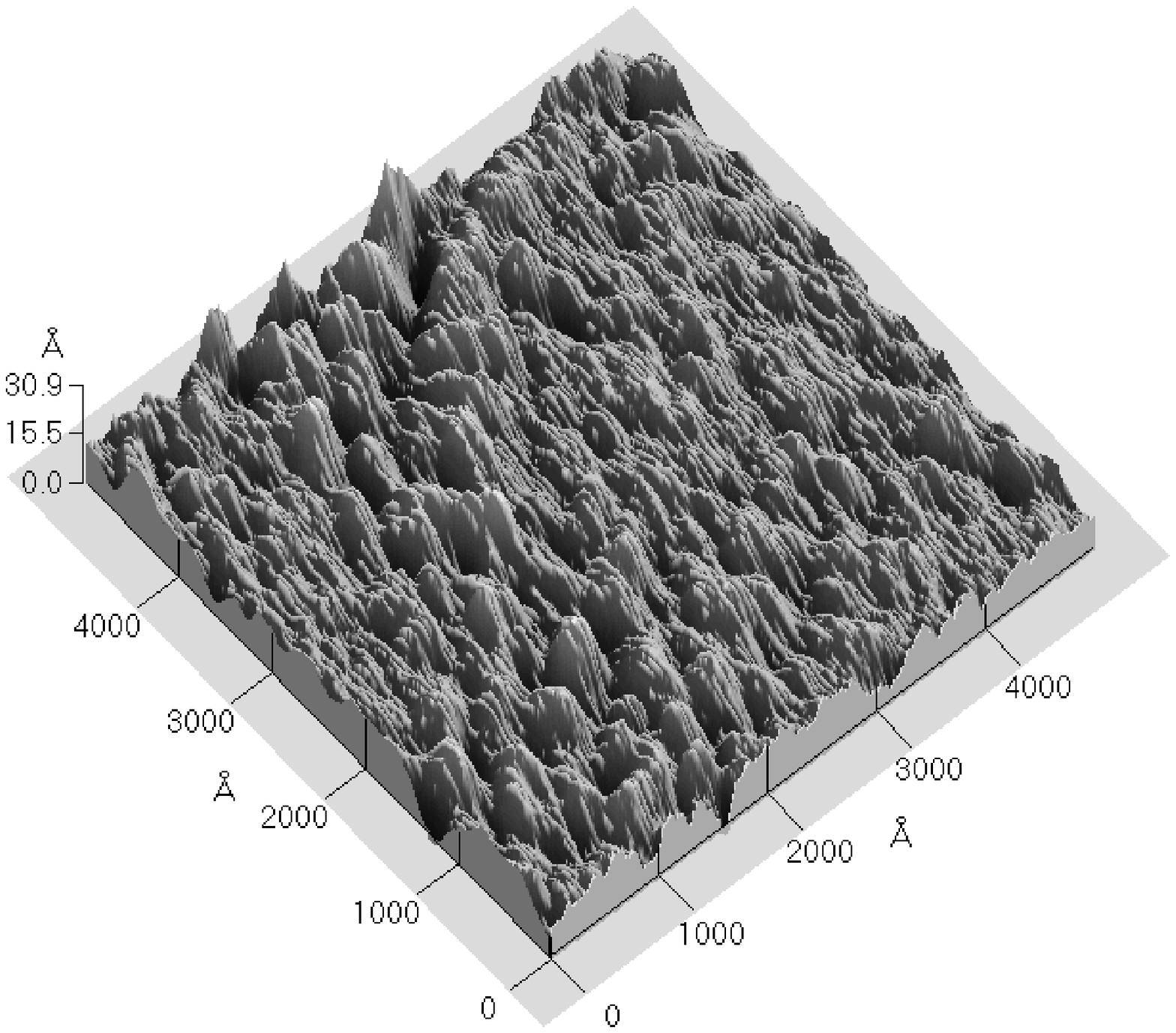}
\epsfxsize=7.4truecm\epsfbox{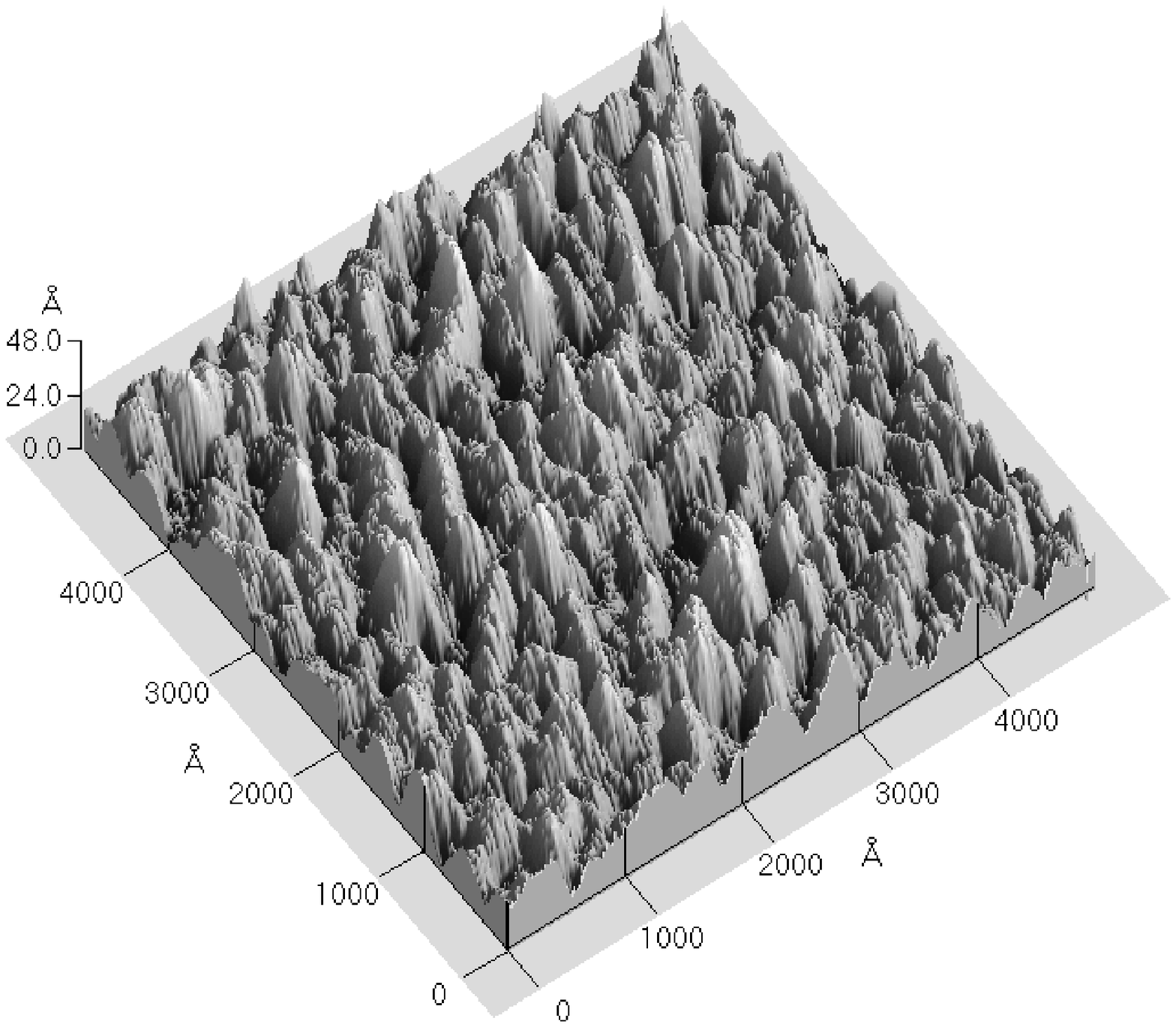}
\epsfxsize=7.4truecm\epsfbox{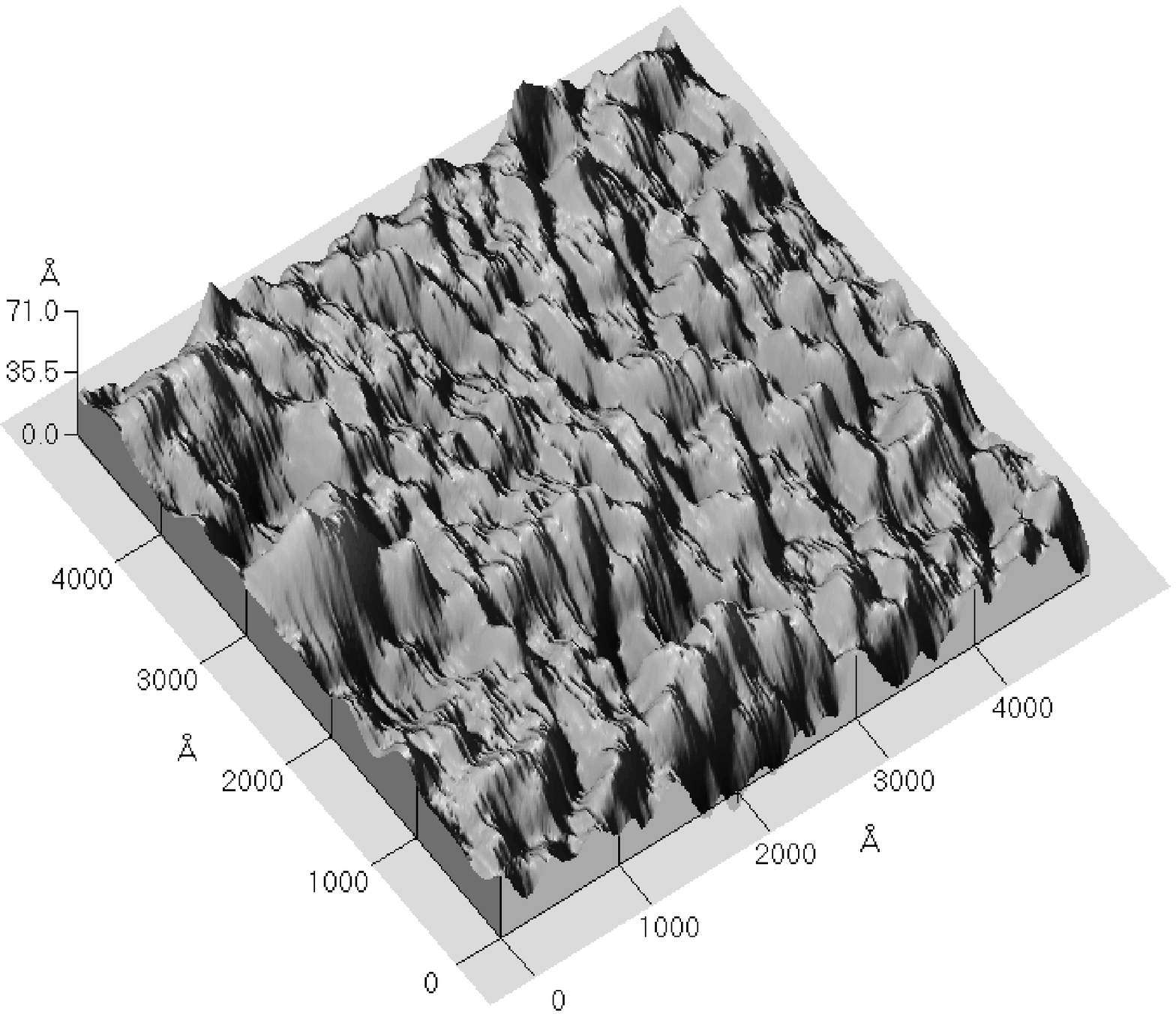}

\narrowtext \caption{ AFM surface images  (all $ 0.5\times 0.5 \mu
m^2 $ ) of $V_2 O_5$ films with thickness of  (a) 20nm , (b)100nm, (c)
260 nm. (from up to down) }
\end{figure}
The quantitative information of the surface morphology can be
derived by considering a sample of size $L$, and defining the mean
height of growing film $\bar{h}$ and, its roughness $w$  through
[10]:
 \be \bar{h} (L,t) = \frac{1}{L} \int_{-L/2} ^{L/2}  dx
h(x,t)
\ee
 and
 \be
  w(L,t) =( \langle ( h - \bar{h})^2 \rangle
)^{1/2}.
 \ee
$\langle \cdots \rangle$ denotes an averaging over different
realization (samples). Starting from a flat interface (one of the
possible initial conditions), it was conjectured by Family and
Vicsek [11] that a scaling of space by factor $b$ and of time by
a factor $b^z$ ($z$ is the dynamical scaling exponent), re-scales
the roughness $w$ by factor $b^{\chi}$ as follows: $w( bL, b^zt)
= b^{\chi} w( L,t)$, which implies that \be w(L,t) = L^{\chi}
f(t/L^z). \ee If for large $t$ and fixed $L$ $(t / L^z
\rightarrow \infty)$, $w$ saturates then $f(x) \rightarrow
const.$, as $ x \rightarrow \infty$. However, for a fixed large
$L$ and $1<< t << L^z$, one expects that correlations of the
height fluctuations are set up only within a distance $t^{1/z}$
and thus must be independent of $L$. This implies that for $x <<
1$, $f(x)\sim x^{\beta}$ with $\beta=\chi / z$. Thus dynamic
scaling postulates that, $w(L,t) \sim t^{\beta}$ for  $ 1 << t <<
L^z $ and $\sim L^{\chi}$ for $ t
>> L^z$. The roughness exponent $\chi$ and the dynamic exponent
$z$ characterize the self-affine geometry of the surface and its
dynamics, respectively. We measure the exponent $\chi$ from
equal-time height-height correlation function defined as $C_2
(l=|x_1 -x_2|) = < |h(x_1) - h(x_2) |^2 >$. Here $h(\bf{x})$ is
the surface height at position $\bf x$ on the surface relative to
the mean surface height.

 In order to determine the roughness exponent $\chi=\xi_2/2$,  we
 consider the
roughness of the samples with thickness $20 nm$, $100 nm $ and
$260 nm$. Fig.3 shows that, for the thickness $ 20 nm$ and $100
nm$, the scaling behavior exists only for the scaling region
$\sim 4 nm$ to $20 nm$, but for the stationary state sample, $<h>
= 260 nm$, there is a cross over for the scaling exponent
$\chi=\xi_2/2$ in $ l = l* \approx 22nm$. The exponent is
${\xi_2} _1$ for the length scales $4 nm$ to the scale  $l*$, and
for the length scales $ 25 nm \leq l \leq 50 nm$, the fluctuation
is determined by another exponent ${\xi_2} _2$. The measured
values for the exponents ${\xi_2} _1$ and $ {\xi_2} _2$ are $1.64
\pm 0.06$ ($\chi_1 = 0.83\pm 0.03$) and $0.58 \pm 0.08$ ( $\chi_2
= 0.29\pm 0.04$) respectively. The measured roughness exponent
$\chi$ of the samples $20nm$, $100nm$ and $150nm$ are $0.71 \pm
0.04$, $ 0.77 \pm 0.03$ and $0.82 \pm 0.03$, respectively.
As shown in Figs. 4 and 5, we note that the correlation length $l*$ is about $22 nm$.
Therefore, we are dealing with the statistical properties of the correlated scaling surfaces.
As discussed in [22], such systems exhibit the sampling-induced hidden cycles
(log-periodic fluctuations [23]).
Such oscillatory behavior will diminish when the sampling size is sufficiently large.
The oscilation amplitude approaches zero to within an order of $ \delta = \sqrt(l*/{M L})$, where $L$ and
$M$ are the sampling size and the number of independent curves which are averaged.
We have determined each exponents by averaging on the eight AFM images.
In our averaging it appears that there is no log-perodic property for
the height-height correlation function. Therefore the surfaces of $V_2 O_5$ are
self-affine.

 The existence of the cross over scale
$l*$ in the stationary state is not observed only in the second
moment $C_2$. In Fig.4 log-log plot of $C_3$ vs $l$ for the three
samples are presented. It is evident that for the stationary
state sample, the scaling exponent of the third moment $C_3$ has
also a cross over in $l*$. The measurement shows that in the
stationary state the exponents $  {\xi_3}_i$ behave as ${\xi_3}_i
=3/2 {\xi_2}_i $ with $i=1,2$.

 We also
examine the scaling behaviour of the q-th moment of the
height-difference $C_q = <|h(x_1) - h(x_2)|^q >$ and show that
all of the moments at least up to $q=20$ behave as  $|x_1 -
x_2|^{\xi_q} $. Fig.5 shows $\xi_q$ vs $q$ for $260nm$ sample in
 $ 25 nm \leq l \leq 50 nm$ scaling region. The graph
of $\xi_q$ for the $ 4 nm \leq l \leq 20 nm$ region is the same as
Fig.5 but with different slope. In the two scaling regions, the
$\xi_q$'s have a linear dependence on $q$. This measurement
indicates that the height-fluctuations are not intermittent and
all of scaling exponents in the stationary state can be expressed
by ${\xi_2}_i$, $i=1,2$ only.

\begin{figure}
\epsfxsize=9truecm \epsfbox{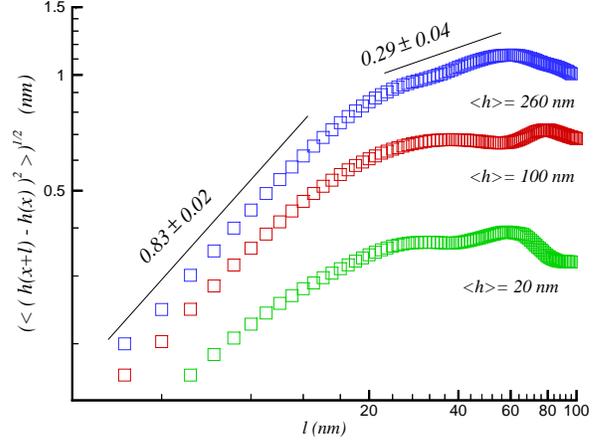} \narrowtext \caption{
Log-log plot of the second moment of height-difference vs $l$,
which shows that for samples with thickness 20nm and 100nm the
scaling region is $ 4nm < l < 20nm $ and for the sample with
thickness $260nm$ there is a cross over scale  $l* \approx 22 nm$.
The upper and lower parts of height-height correlation function
have the roughness exponent $0.29 \pm 0.04$ and $0.83\pm 0.03$,
respectively.
 }
\end{figure}

\begin{figure}
\epsfxsize=9truecm \epsfbox{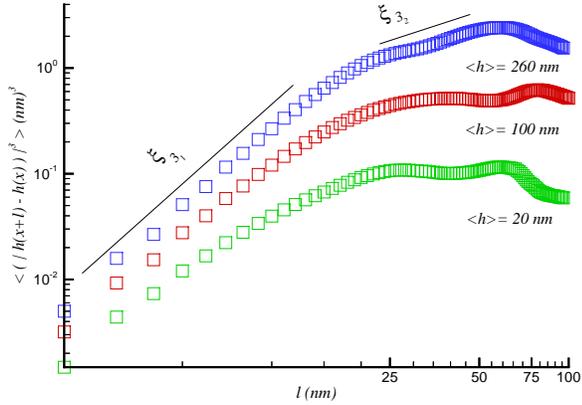} \narrowtext \caption{
log-log plot of the third moment of height-difference vs $l$,
which shows that in the length scale $l* \approx 22 nm$ there is
also a cross over for the scaling exponent $\xi_3$ of the sample
$260 nm$. }
\end{figure}

\begin{figure}
\epsfxsize=7.9truecm \epsfbox{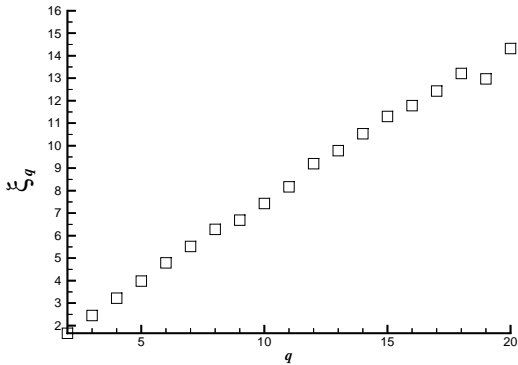} \narrowtext \caption{The
plot of $\xi_q$ vs $q$, which has a linear dependence on q and
shows that the height-difference fluctuations are not
intermittent.}
\end{figure}
There is another method to check that the intermittency is absent
in the height fluctuations of $V_2 O_5$ surface [12-13]. Let's
introduce the generating function $Z(q,N;l)$. The generating
function is defined through \be Z(q,N;l) = \sum_{i=1} ^N \mu_i^q,
\ee where the normalized measure
 $ \mu_i \geq 0$ is
\begin{equation}
 \mu_i = \frac{ |h(x_i + l) - h(x_i)|}{ \sum_{i=1}
^N |h(x_i + l) - h(x_i)|},
\end{equation}
 and, $N$ is the number of
data $|h(x_i + l) - h(x_i)|$. For large $N$, the generating
function $Z$ scales as
 $ Z(q,N;l)  \sim N^{- \tau(q)}$.
We measure the exponent $ \tau(q) = (q-1) D_q$ in both scaling
region ( $ 4nm$ to $20nm$ and $25nm$ to $50nm$).
 The amplitude of $Z$ depends only on $l$ and $q$.
In Fig.6 we plot the $(1-q) D_q$ vs $q$ for small $ l $.
  The figure shows that the $ \tau(q) $ has a linear
dependence on $q$ and the generalized fractal dimension $D_q$ is
independent of $q$. Its value is $D_q = D_0 = 7.34 \pm 0.02 $.
Therefore we conclude that the height fluctuation is not
intermittent.
 It is necessary to note that the exponents $ \xi_q$ and $D_q$
 are not independent in the small region of $l$.  Defining $\xi_q = q H_q$, it has been
shown in [20] that $H_q$ can be expressed in terms of $ D_q$ as;
\be H_q= H_1 + \frac{(q-1)(D_q - D_{eff})}{q}.
 \ee
Our measurement for $H_q$ shows that the $H_q$ is independent of
$q$ and this means that $D_{eff} = D_q = D_0$. Let us discuss the
origin of the $D_{eff}$ in eq.(6). To evaluate the $C_q$ we
should calculate the summation $\frac{1}{N} \sum_{i=1} ^{N} |
h(x_i) - h(x_i + l) |^q$, where $N$ is the number of points over
which the average is taken. Normally small $l$ is meant by $l
\sim 1/N $. The authors of Ref.[20] assumed that, in evaluation of
 $C_q$, $l$ and $N$ may be related in a way different from  $l
\sim 1/N $. That is $N \sim l ^{- D_{eff}}$ ( $D_{eff}$ could be
considered here as a fractal dimension of the effective support
of the process). The choice of a particular partition has no
effect on the $H_q$ spectrum. However, $D_{eff}$ enters the
relation between $H_q$ and generalized dimension $D_q$ as
expressed in eq.(6).

\begin{figure}
\epsfxsize=7.9truecm \epsfbox{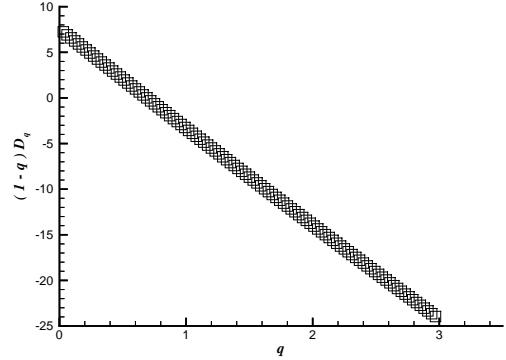} \narrowtext \caption{ Plot
of $(1-q) D_q$ vs $q$, shows that the fluctuations are determined
only by dimension $D_0$ and the surface are not intermittent in
the scaling region.}
\end{figure}

\section{Discussion and Growth model }

In our experiments we obtain a region in which the growth
exponents of $V_2 O_5$ are consistent with early-scaling regime of
forced Kuramoto-Sivashinsky (KS) equation in 2+1 dimensions [19].
 The forced
Kuramoto-Sivashinsky has the following form [14-19],

\be
 \frac{\partial h}{\partial t} =   \nu \nabla^2 h - k \nabla^4 h
 +
 (\lambda /
2) ( \nabla h)^2 +  \eta(x,t) \ee
 where the  $\nu$,$k$ and $
\lambda$ are the surface tension, surface diffusion   and
non-linear factor, respectively. The force $\eta$ is a noise term
reflecting spatial and temporal fluctuation in the incoming flux
of material and has a Gaussian distribution and uncorrelated in
space and time. In the limit $k=0$ and for $\nu > 0$ the forced
KS equation reduces to the Kardar-Parisi-Zhang (KPZ) equation.
The KS equation with $\lambda = 0$ is a linear equation and can be exactly solved by the standard methods [19,22]. For some Gaussian noise term, because of the linearity the
probability distribution function (PDF) of $h-\bar h$ is also Gaussian.
It is discussed in [22] that for $\lambda =0$ and $\nu/k < 0$ the KS equation generates a mound
surface and for $\nu/k >  0$ gives a self-afine surface, with a roughness exponent $\simeq 1$.
For $\lambda \neq 0$
the nonlinear term breaks the symmetry under transformation $ h \rightarrow -h$.
Therefore the PDF of $h-\bar h$ must be skewed. Our measured value for the PDF of
$h-\bar h$ shows that the PDF has positive skewness.
We will report the skewness and kurtosis of the PDF of the $V_2 O_5$ films elsewhere [12].

Recent simulation of KS equation reveals the presence of the early
and long scaling regimes [19]. The initial-time values for the
growth exponent $\beta$ and the roughness exponent $\chi$ are
found to be $0.22 - 0.25$ and $0.75 - 0.80$ respectively. The
long time scaling regime is determined by the exponents $\beta=
0.16-0.21$ and $\chi=0.25-0.28$. The scaling exponents are
notably less than the exponents of KPZ equation [19].
 The long-time behavior of the KS-equation has an interesting feature.
 For long-times the height-height correlation
function exhibits a bifractal structure with two different
roughness exponents (see Fig.11 in ref. [19]). For $\lambda =
2.0$, $ \nu = -0.2$ and $ k=2.0$, the upper and lower parts of
height-height correlation function have the roughness exponent
$0.27$ and $0.71$, respectively.

Now let us compare our results with the numerical simulation of
KS equation. Using the numerical results of ref.[19] in table II,
one can observe that for early-scaling regime of the
Kuramoto-Sivashinsky equation the exponents $ \chi$ and $ z$
satisfy the exponent-identity $ \chi + z \simeq 4$. Using the
time dependence of roughness of the samples with thickness $
20nm$ , $ 50nm$, $ 100nm$, $120nm$ and $150nm$ we find that
$\beta =0.29 \pm 0.04 $ (see Fig.7) and therefore $ \chi + z =
3.7 \pm 0.4$, which satisfies the exponent-identity within the
experimental errors. We also observe the similar bifractal
structure of the height-height correlation function in our
samples with thickness $260nm$ (see  Fig.11 in [19]). To relate
the cross over scale $l*$ to the the surface morphology, we
have measured average distance of nearest local maxima $\bar d_{max}$,
 and obtain that the  $\bar d_{max}$ is
in the same order of magnitude of $l*$. This means that, in the
stationary state, in average, if the distance between the points
$x_1$ and $ x_2$ lies between the two local maxima of $h(x)$, the
dynamics is determined by the roughness exponent $\chi_1$. Also
for the points that the relative distance lies between the next
neighbor maximums ( $\sim 20-40 nm$ ), the dynamics is given by
another exponent $\chi_2$ which is less than the KPZ roughness
exponent. Our measurements show that the roughness exponent
$\chi_2 = 0.29 \pm 0.04$ is less than the the KPZ roughness
exponent in $2+1$ dimensions i.e. $ 0.38$.


\begin{figure}
\epsfxsize=7.9truecm \epsfbox{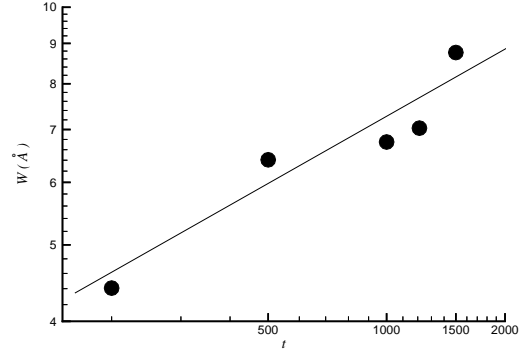} \narrowtext \caption{
Log-log plot of the interface width $w$ vs deposition time or
thickness (in \AA) $t$ for the samples with thickness $20nm$,
$50nm$, $100nm$, $120nm$ and $150nm$. The measured value for
$\beta $ is $0.29 \pm 0.04 $.}
\end{figure}

\begin{figure}
\epsfxsize=7.9truecm \epsfbox{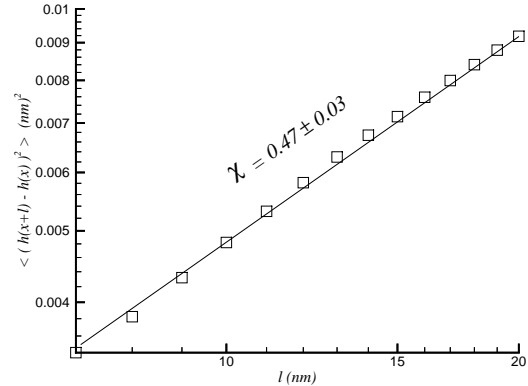} \narrowtext \caption{
Log-log plot of the second moment of height-difference vs $l$,
for polished Si(100) substrate. The measured value for the
roughness exponent is $\chi_{Si}= 0.47 \pm 0.03$. }
\end{figure}

\begin{figure}
\epsfxsize=7.4truecm\epsfbox{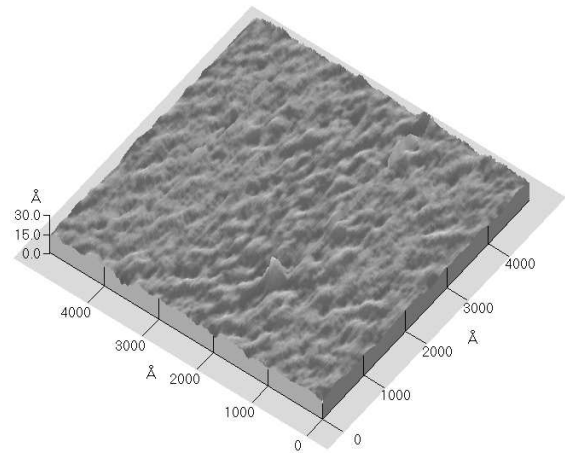}

\narrowtext \caption{ AFM image of polished $Si(100)$ substrate.}
\end{figure}

In summary we have checked the intermittency of height fluctuation
of $V_2 O_5$ via two different methods. These methods predict that
the statistics of height fluctuations is not intermittent. We
measure the scaling exponents of height-difference moments, $C_q (
l =|x_1 -x_2|)$, and show that for  small $l$, they behave as $
\sim |x_1 - x_2 |^ {\xi _q}$.
 The obtained $\xi_q$`s are linear function of $q$.
The observation can be explained with the recent numerical
simulation on the basis of nonlinear KS equation. However, one
should note that for very early stages, when $\nu$ term dominates
$k$ term, instability arises in x or y direction causing the
ripple structure with corresponding wave vector in x or y
direction in KS equation. When the ripple structure is formed,
there exists slope asymmetry that activates the nonlinear effect.
Therefore, in the early-stage of the process, the ripple structure
transforms to symmetric mounds. In our experiments we were not
able to detect the ripple structure for the samples with $<h>$
$\ll$ $20nm$. It could mean that the polished $Si(100)$
 substrate may have an initial roughness which may destroy the ripple structure of the
 film for very early-stages of the growth. We measure the roughness
 exponent of the silicon substrate and find that the $\chi_{si} = 0.47\pm
 0.03$ (see Figs.8 and 9).

 Finally we note that the measured
roughness exponents $\chi$ and $\beta$ can be higher than the
true values because of the tip effect [21]. Aue etal. showed that
the surface fractal dimension (fractal dimension $d_f$ for 2+1
interface is related to $\chi$ by $d_f= 3 - \chi$ )
 determined with a scanning probe
technique will always lead to an underestimate of the actual
fractal dimension, due to the convolutions of tip and surface.
Their analysis included tips with different shapes and aspect
ratios. For a tip similar to what we have used, it is suggested
that our true $\chi_1$ and $\chi_2$  should be around $\sim 0.75$
and $\sim 0.25$, respectively. We note that the corrected
exponents also satisfy the exponent identity.

{\bf Acknowledgement}\\

We thank A. Aghamohammadi, F. Azami, M.M. Ahadian,  J. Davoudi, A.
Farahzadi, G. Ketabi and Z. Vashaei
 for useful discussions.

\end{document}